\documentclass[a4paper]{jpconf}
\usepackage{iopams}
\usepackage{graphicx}
\usepackage{subfigure}

\makeatletter
\DeclareRobustCommand{\subref}[1]{%
  \ifx|#1|% 
    \protect\G@refundefinedtrue%
    \@latex@warning{\string\subref\space has no argument}
  \else\expandafter\@subref\csname r@#1\endcsname{#1}\fi} 
\def\@subref#1#2{%
  \ifx#1\relax%
    \protect\G@refundefinedtrue
    \nfss@text{\reset@font\bfseries ??}%
    \@latex@warning{Reference `#2' on page \thepage\space undefined}%
  \else% 
    \xdef\r@subref{\expandafter\@firstoftwo#1}%
    \expandafter\@@subref\r@subref%
  \fi%
}
\def\@@subref#1(#2){(#2)}
\makeatother

\newcommand\Nch{\ensuremath{\mathrm{N_{ch}}}}
\newcommand\dNdeta{\ensuremath{\mathrm{d\Nch{}/d\eta}}}
\newcommand\vn[1]{\ensuremath{\mathrm{v_{#1}}}}

\newcommand\vncum[2]{\ensuremath{\mathrm{\vn{#1}\{#2\}}}}

\begin{document}
\title{Charged-particle pseudorapidity density and anisotropic flow
  over a wide pseudorapidity range using ALICE at
  the LHC} \author{K. Gulbrandsen$^1$ for the ALICE Collaboration}
\address{$^1$ Niels Bohr Institute, University of Copenhagen,
  Blegdamsvej 17, 2100 Copenhagen \O, Denmark} \ead{gulbrand@nbi.dk}

\begin{abstract}
The pseudorapidity density and anisotropic flow of charged-particles
provide fundamental information about global variables and
correlations in heavy-ion collisions. The pseudorapidity density is
related to the energy available for particle production while the
anisotropic flow is related to collective effects from interactions
between these particles. Extending these measurements to very forward
pseudorapidities yields information about
the longitudinal expansion of the system. The first measurements
performed at the LHC over more than 8 units of pseudorapidity are
presented. The longitudinal scaling of the measurements is analyzed
and comparison to models is performed.
\end{abstract}

\section*{Introduction}
The goal of colliding heavy-ions is to
understand the strong interaction in the hot dense state known as
the Quark--Gluon Plasma (QGP). One of the most fundamental
characteristics of these collisions is the energy available for
particle production, which can be assessed by measuring the
charged-particle pseudorapidity density, \dNdeta{}. Further
insights can be gained by investigating quantities
which probe the collective behavior of the produced particles.
Anisotropic flow \cite{PhysRevD.46.229} is one signature of this
collectivity arising from multiple interactions between particles
which transfers the initial spatial anisotropy of the collision into
a momentum anisotropy of the produced particles. It is
quantified as the measurement of the flow harmonics, \vn{n}, of
the particle distribution \cite{Voloshin:1994mz}.
Extending the measurements to forward pseudorapidity
allows for an assessment of the physics in the fragmentation region
and clarifies the role of fluctuations in the initial energy in the
anisotropic flow measurement.

\section*{Detector and data analysis}
The measurements are carried out using the ALICE detector
\cite{1748-0221-3-08-S08002} at CERN. Three main sub-detectors, each with full
($2 \pi$) azimuthal coverage, are utilized in these measurements. At
mid-rapidity, the Silicon Pixel Detector (SPD) \cite{1748-0221-3-08-S08002} is
used. This detector has 2 cylindrical layers of silicon at radii of 3.9 and 7.6 cm
from the beampipe with pseudorapidity coverages of $|\eta| < 2$ and $|\eta| < 1.4$,
respectively. With almost 10 million total channels, the SPD
provides excellent spatial resolution in both $\eta$ and $\varphi$ for
registering particles produced in the collision. The detector's close proximity
to the beampipe reduces the number of detected secondary particles.

At forward rapidities the Forward Multiplicity Detector (FMD)
\cite{Christensen:2007yc} and the VZERO \cite{forward_tdr} are used. The FMD is
divided into three sections at different positions along the beam axis with each
section made of one or two rings of silicon strip sensors. Each ring has 10240
silicon strips with high $\eta$ resolution and 20 or 40 azimuthal sectors for
resolving $\varphi$. The nominal acceptance of the FMD is $-3.4 < \eta < -1.7$
and $1.7 < \eta < 5.0$. The VZERO has a nominal acceptance of $-3.7 < \eta < -1.7$
and $2.8 < \eta < 5.1$. This detector is composed of scintillator counters whose
signal is proportional to the number of particles incident upon the detector.
The detector has 4 radial segments for $\eta$ resolution and 8 azimuthal segments
for $\varphi$ resolution.

\begin{figure*}[htbp]
  \centering
  \includegraphics[keepaspectratio,width=\textwidth]{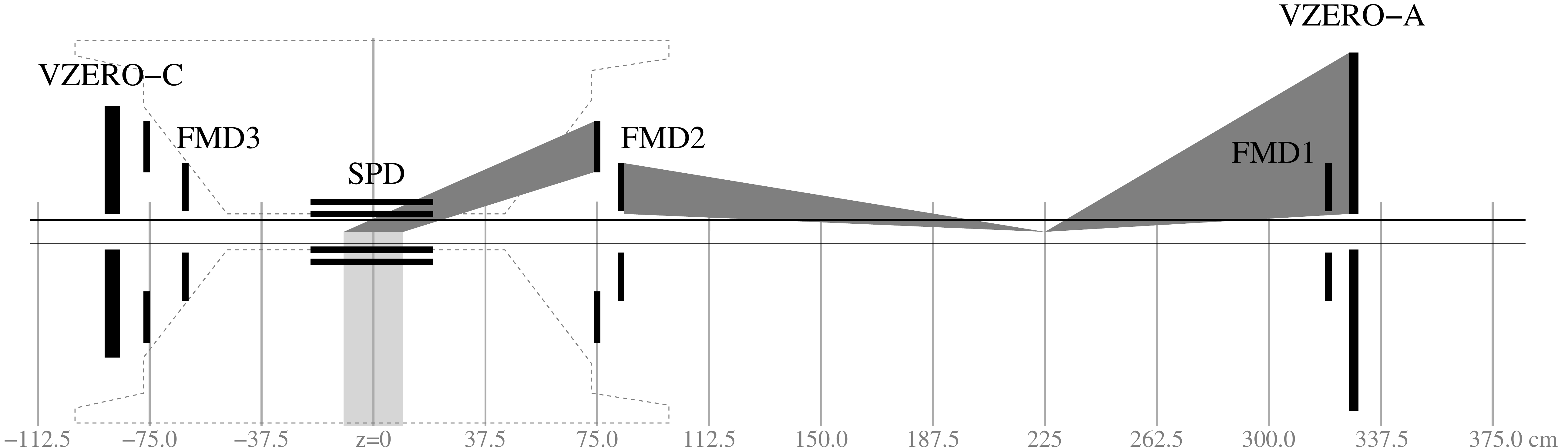}
  \caption{Drawing of the cross-section of the sub-detectors used for
    this analysis and locations of the nominal ($z = 0$) and `satellite'
    interaction points. The dashed line indicates where dense support
    material exists.}
  \label{layout}
\end{figure*}

Figure~\ref{layout} shows a schematic cross-section of these detectors
and the trajectories which particles would follow from different interaction
vertices. From the nominal vertex at $z = 0$, few secondary particles
are detected in the SPD, but for FMD2, FMD3, and VZERO-C many secondary
particles produced in support material are detected. This effect
requires a very accurate material description and a correction dependent
on simulations. An alternate method was, therefore, developed to measure
\dNdeta{} where only FMD2, FMD1, and VZERO-A were used to
extend the mid-rapidity measurement of \dNdeta{} from the
SPD. `Satellite' vertices spaced in intervals of 37.5~cm from $z = 0$
created via beam debunching during injection and acceleration \cite{LHC_sat}
were used to avoid large corrections due to secondary particle production
and also to extend the $\eta$ coverage for the \dNdeta{} measurement to
$-5.0 < \eta < 5.5$. The collision rate with displaced vertices is about
1000 times smaller than at the nominal vertex position which does not
allow for a precision measurement of \vn{n} with this method.

\section*{Pseudorapidity density}
\begin{figure}[htbp]
  \centering
  \begin{tabular}{@{}p{.4375\textwidth}@{}p{.5625\textwidth}@{}}
    \subfigure[]{%
      \includegraphics[keepaspectratio,width=.4375\textwidth]{%
        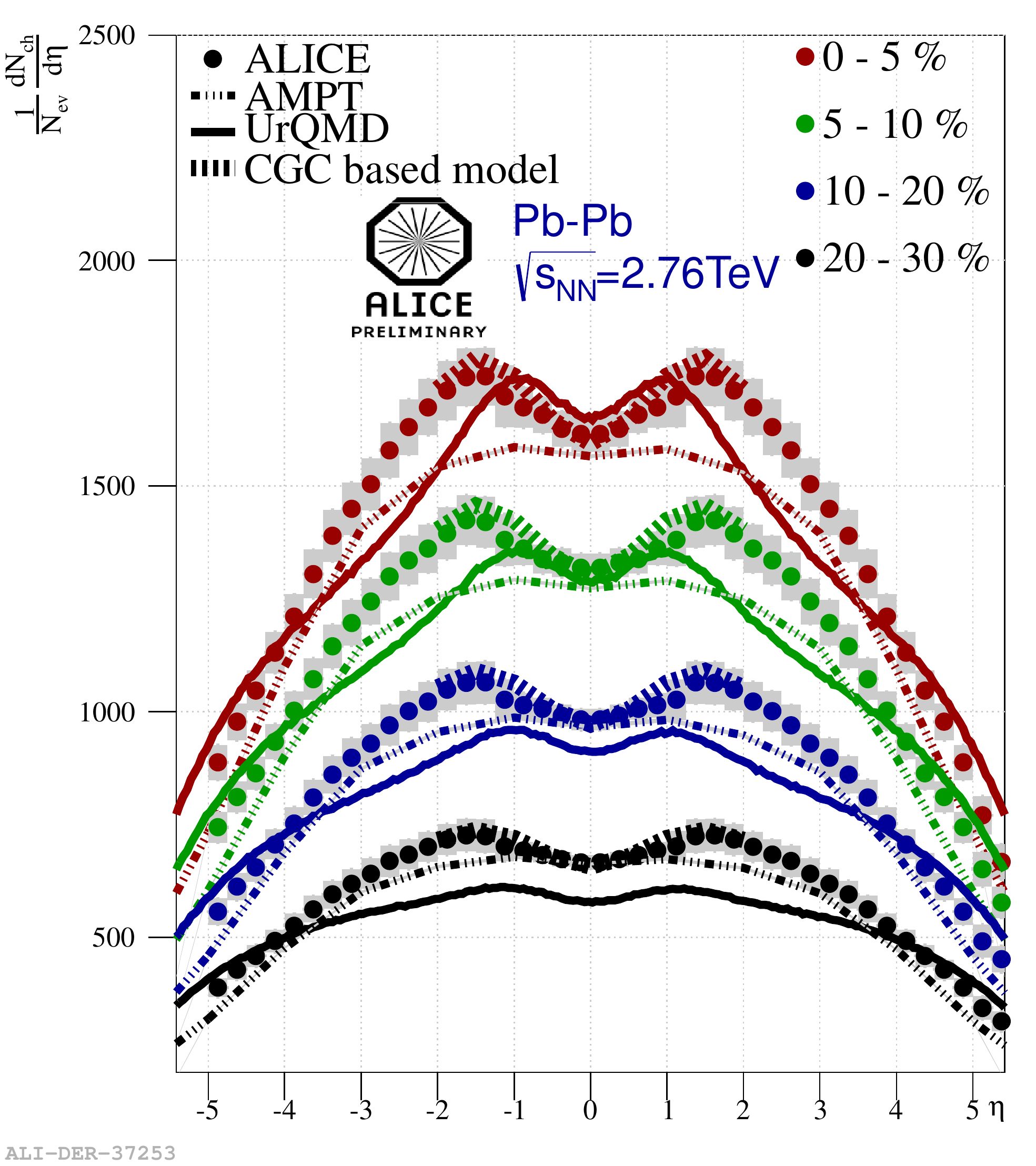}
      \label{dndetaresult}}
    &
    \subfigure[]{%
      \includegraphics[keepaspectratio,width=.5625\textwidth]{%
        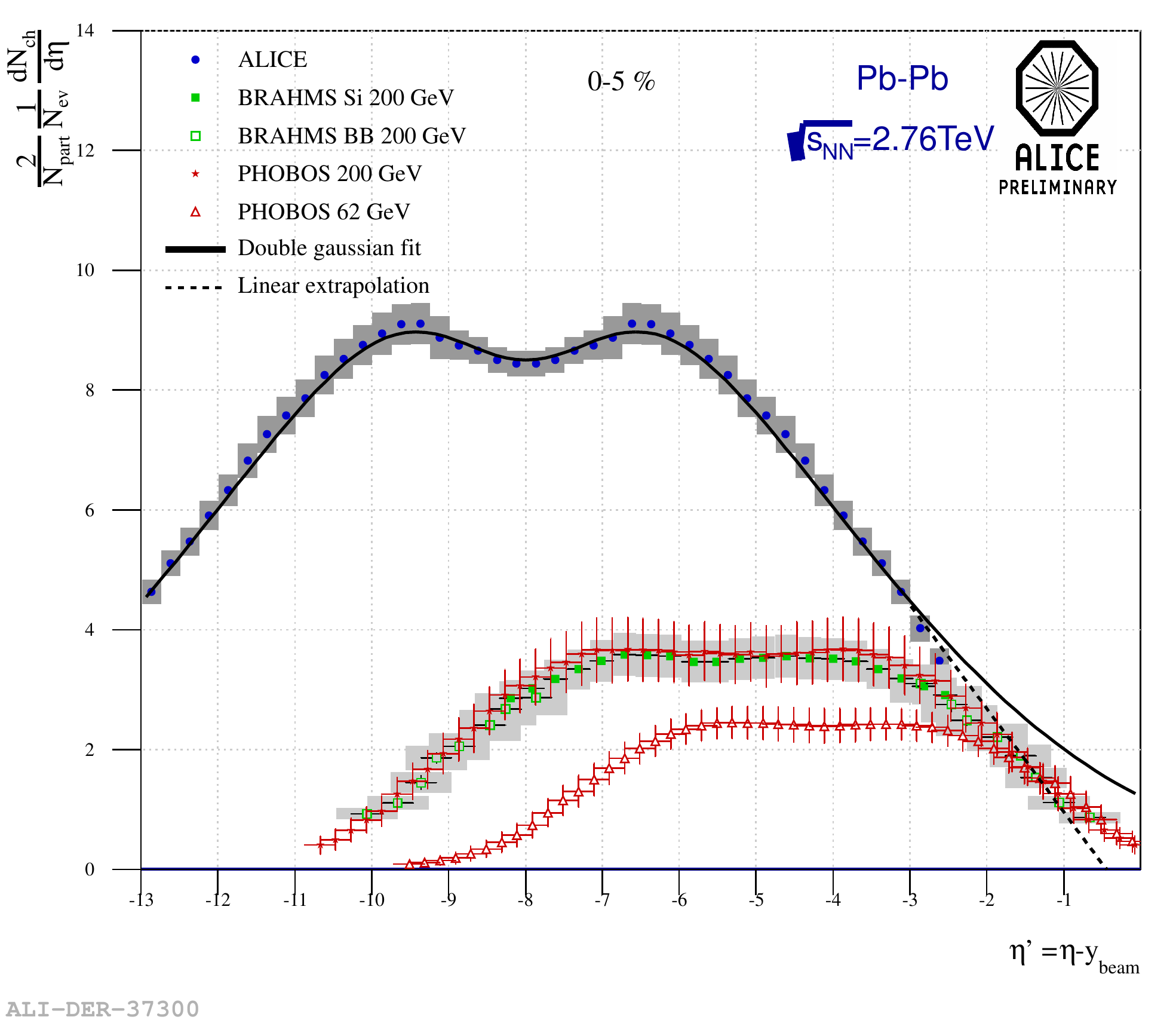}
      \label{dndetalongscal}}      
  \end{tabular}
  \caption{\subref{dndetaresult} \dNdeta{} versus $\eta$
    measured by ALICE in Pb--Pb collisions at $\sqrt{s_{NN}} = 2.76$ TeV
    in 4 different centrality bins. The data are symmetrized in the
    region of $|\eta| < 5$ to reduce systematic errors. Model
    comparisons are shown
    \cite{PhysRevC.72.064901,PhysRevC.83.034904,Mitrovski:2008hb,ALbacete:2010ad,Albacete:2011fw}.
    \subref{dndetalongscal} \dNdeta{} per participant pair versus
    $\eta - \mathrm{y_{beam}}$. The most central ALICE data are compared to lower
    energy measurements \cite{PhysRevC.83.024913,PhysRevLett.88.202301}. The
    full drawn line is an extrapolation of the distribution from a fit to the
    difference of two Gaussian distributions centered at $\eta = 0$, but with
    different widths. The dashed line is a straight line fit to the forward
    pseudorapidity points of the measured distribution.}
\end{figure}

Figure~\ref{dndetaresult} shows the result of the \dNdeta{}
analysis using the SPD with nominal vertex collisions for the
region of $|\eta| < 2$ and the FMD and VZERO detectors with `satellite'
vertex collisions. The results are compared to three models. The AMPT
model \cite{PhysRevC.72.064901} tuned to the measured mid-rapidity
\dNdeta{} \cite{PhysRevC.83.034904} fails to reproduce the shape of the \dNdeta{}
distribution. UrQMD \cite{Mitrovski:2008hb} predicts the \dNdeta{} shape in the
region of $|\eta| < 1$ in the two most central bins, but does not properly describe
the distribution at forward $\eta$. The CGC based model
\cite{ALbacete:2010ad,Albacete:2011fw} reproduces the \dNdeta{} shape within the
limited pseudorapidity range of the prediction, $|\eta| < 2$.

Figure~\ref{dndetalongscal} shows the distribution of \dNdeta{} per participant
pair as a function of $\eta - \mathrm{y_{beam}}$. The most central ALICE data are
compared to lower energy data from RHIC
\cite{PhysRevC.83.024913,PhysRevLett.88.202301}. Near beam rapidity the lower
energy distributions overlap, suggesting the validity of longitudinal scaling in
the fragmentation region \cite{PhysRev.188.2159,PhysRevLett.91.052303}.
The ALICE acceptance does not extend far enough to show an overlap with
the lower energy data, but extrapolations are consistent with the preservation
of longitudinal scaling in the fragmentation region.

\section*{Anisotropic flow}
\begin{figure}[htbp]
  \centering
  \begin{tabular}{@{}p{.5\textwidth}@{}p{.5\textwidth}@{}}
    \subfigure[]{%
      \includegraphics[keepaspectratio,width=.5\textwidth]{%
        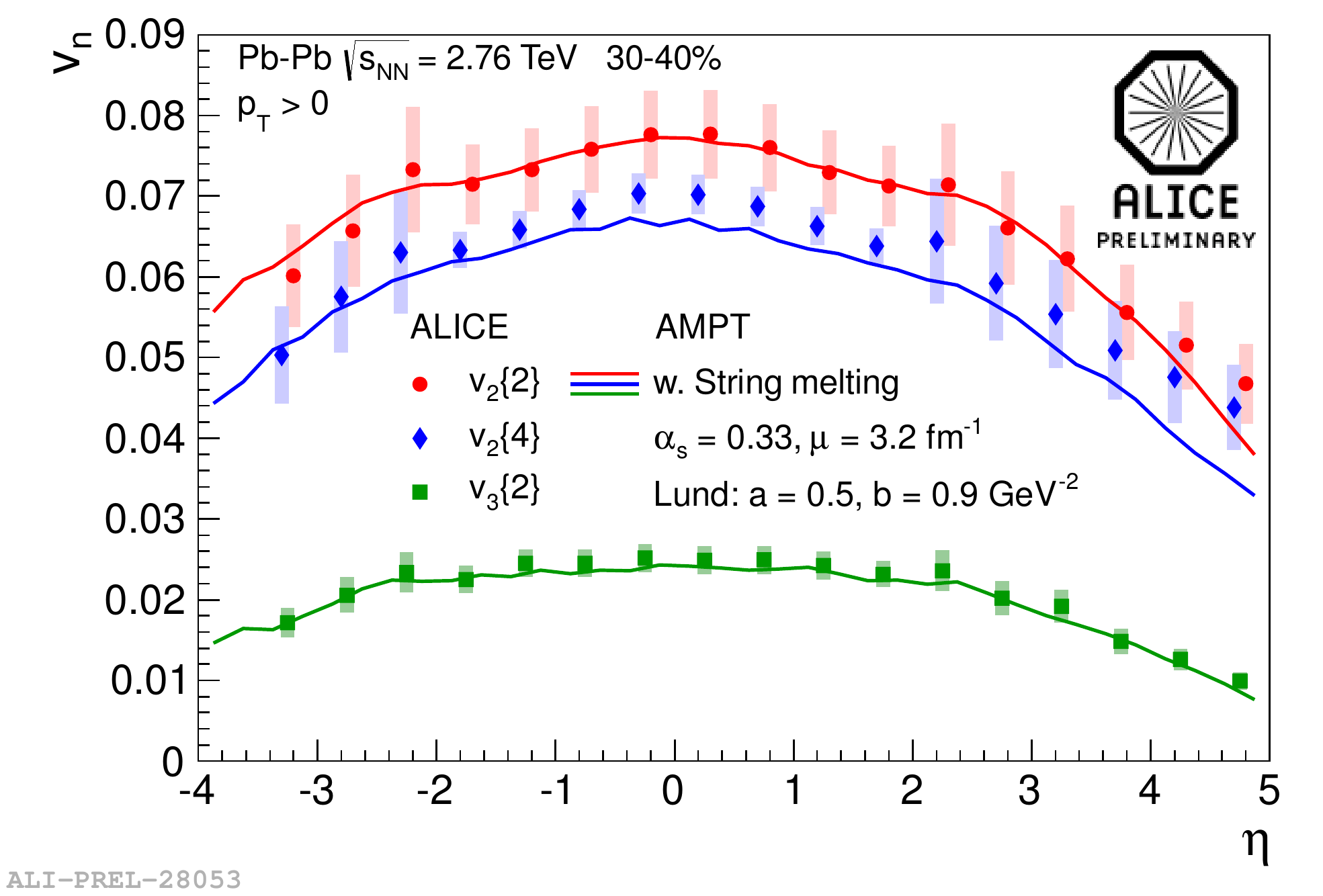}
      \label{vnmodels}}
    & \subfigure[]{%
      \includegraphics[keepaspectratio,width=.5\textwidth]{%
        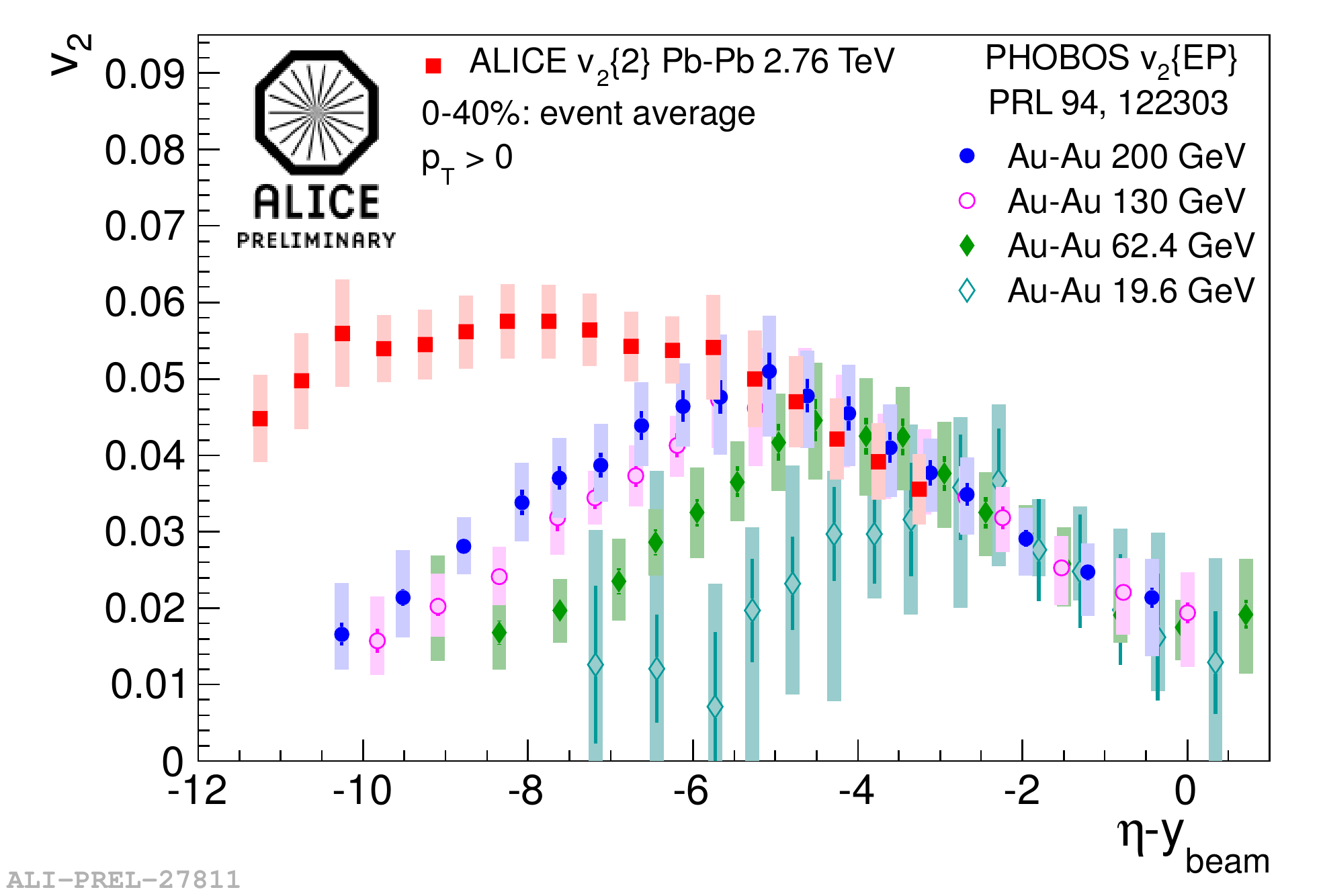}
      \label{v2longscal}}
  \end{tabular}
  \caption{\subref{vnmodels} \vn{2} and \vn{3} versus $\eta$ for the 30--40\%
    centrality class measured with 2 and 4 particle cumulants
    \cite{Bilandzic:2010jr}. The data are compared to the AMPT model
    \cite{PhysRevC.72.064901} with the settings shown \cite{PhysRevC.83.034904}.
    \subref{v2longscal} The measurement of \vn{2} versus
    $\eta - \mathrm{y_{beam}}$. The data are compared to lower energy results
    \cite{Back:2004zg}.}
\end{figure}

The data were analyzed using the cumulant method \cite{Bilandzic:2010jr} to
extract the values of the flow harmonics, \vn{n}. Different multiparticle
correlators were used to understand the effect of flow fluctuations. The analysis
was carried out using only collisions occurring around the nominal vertex region
($|z| < 10$ cm). The inner layer of the SPD was used for the measurement in the
range $|\eta| < 2$ and the FMD (all three sections) extended this measurement to
the range $-3.4 < \eta < 5.0$. Figure~\ref{vnmodels} shows the result of the
analysis. The large value of \vn{2} is due to the harmonic's sensitivity to the
large anisotropy in the initial geometrical shape present in the 30--40\% most
central collisions. The smaller value of \vn{3} is attributed to the harmonic's
dominant sensitivity to fluctuations in the initial geometrical shape. The
essentially constant difference between \vncum{2}{2} and \vncum{2}{4} implies that
the effect of flow fluctuations is similar at mid-rapidity and forward rapidity.
The data are compared to the tuned AMPT model \cite{PhysRevC.83.034904}. With this
tuning, the model reproduces the shape of measured flow harmonics as a function of
$\eta$.

Figure~\ref{v2longscal} shows \vn{2} versus $\eta - \mathrm{y_{beam}}$ compared to
lower energy data \cite{Back:2004zg}. Close to beam rapidity,
$|\eta - \mathrm{y_{beam}}| \sim 3$, \vn{2} measured at RHIC and LHC seems to follow
a universal dependence, hinting that longitudinal scaling in the fragmentation region
seen at RHIC energies \cite{Back:2004zg} also holds at the LHC.

\section*{Summary}
The charged-particle pseudorapidity density, elliptic flow (\vn{2}), and
triangular flow (\vn{3}) have been measured over more than 8 units of
pseudorapidity. The available models do not predict the full \dNdeta{} shape. A
large value of \vn{2} is seen for mid-central collisions (30-40\%), consistent
with the anisotropy of the initial geometrical shape of the collision, while
\vn{3} is smaller, consistent with its sensitivity to fluctuations in the shape.
The contribution of flow fluctuations to the flow signal appears to be similar
over a wide range of pseudorapidity. The measured \dNdeta{} and elliptic flow
appear to exhibit longitudinal scaling in the fragmentation region up to
$\sqrt{s_{NN}} = 2.76$ TeV.

\section*{References}

\bibliography{Hot_Quarks_2012_proceedings_Gulbrandsen}

\end{document}